\documentclass[useAMS,usenatbib, usegraphicx]{mn2e}

\newcommand{\kms}{\mbox{km s$^{-1}$}}
\newcommand{\arcdeg}{\mbox{$^\circ$}} 

\newcommand{\muas}{\mbox{$\mu$as}} \newcommand{\muasyr}{\mbox{$\mu$as
    yr$^{-1}$}} 

\newcommand{\Msol}{\mbox{M\raisebox{-.6ex}{$\odot$}}}
\newcommand{\Rsol}{\mbox{R\raisebox{-.6ex}{$\odot$}}}

\newcommand{\Jb}{\mbox{Jy beam$^{-1}$}}
\newcommand{\muJb}{\mbox{$\mu$Jy beam$^{-1}$}}
\newcommand{\RA}[4]{\mbox{${#1}^{\rm h} \; {#2}^{\rm m} \; {#3}\fs{#4}
    $}} \newcommand{\dec}[4]{\mbox{${#1}\arcdeg \; {#2}\arcmin \;
    {#3}\farcs{#4} $}}  % only in math mode
\newcommand{\ergsHz}{\mbox{erg~s$^{-1}$~Hz$^{-1}$}}
\newcommand{\slugcomment}[1]{\date{#1}} % date & version number

\usepackage{graphicx}
\usepackage{amssymb} \usepackage{color} 
\usepackage[modulo]{lineno} %adds line-numbering, useful for co-authors 
\usepackage{fixltx2e} % this fixes ordering of normal figures % and two-column "figure*"
\usepackage{url} % allows you to do \url{xxx.myurlinhere}
\title[Imaging of SN~2011\lowercase{dh}]{Imaging the Expanding Shell of SN~2011\lowercase{dh}}
\author[A. de Witt et al.]
 {A.~de Witt,$^{1}$\thanks{E-mail:alet@hartrao.ac.za}  
 M. F.~Bietenholz,$^{1,2}$
 A.~Kamble,$^{5}$
 A. M.~Soderberg,$^{5}$
 A.~Brunthaler,$^{3}$
 \newauthor
 B.~Zauderer,$^{5}$
 N.~Bartel,$^{2}$ 
 and M. P.~Rupen,$^{4}$ \\
$^{1}$Hartebeesthoek Radio Observatory, PO Box 443, Krugersdorp, 1740, South Africa\\
$^{2}$Department of Physics and Astronomy, York University, Toronto, M3J~1P3, Ontario, Canada\\ 
$^{3}$Max-Planck-Institut f\"ur Radioastronomie, Auf dem H\"ugel 69, 53121 Bonn, Germany\\
$^{4}$National Research Council of Canada, Herzberg Astronomy and Astrophysics Programs, Dominion
Astrophysical Observatory\\
$^{5}$Harvard-Smithsonian Center for Astrophysics, 60 Garden Street, Cambridge, MA 02138, USA}

\begin{document}

\slugcomment{Version 7.0, \today}

\pagerange{\pageref{firstpage}--\pageref{lastpage}}

\maketitle

\label{firstpage}

\begin{abstract}
We report on third epoch VLBI observations of the radio-bright
supernova SN~2011dh located in the nearby (7.8~Mpc) galaxy M51. The
observations took place at $t=453$~d after the explosion and at a
frequency of 8.4~GHz. We obtained a fairly well resolved image of the
shell of SN~2011dh, making it one of only six recent supernovae for
which resolved images of the ejecta are available.
SN~2011dh has a relatively clear shell morphology,
being almost circular in outline, although there may be some
asymmetry in brightness around the ridge.
By fitting a spherical shell model directly to the visibility
measurements we determine the angular radius of SN~2011dh's radio
emission to be $636 \pm 29$~\muas. At a distance of 7.8~Mpc, this
angular radius corresponds to a linear radius of $(7.4 \pm 0.3) \times
10^{16}$~cm and an average expansion velocity since the explosion of
$19000^{+2800}_{-2400}$~\kms.
We combine our VLBI measurements of SN~2011dh's radius with values
determined from the radio spectral energy distribution under the
assumption of a synchrotron-self-absorbed spectrum, and find all the
radii are consistent with a power-law evolution, with $R \sim
t^{0.97\pm0.01}$, implying almost free expansion over the period
$t=4$~d to 453~d.
\end{abstract}


%%%%%%%%%%%%%%%%%%%%%%%%%%%%%%%%%%%%%%%%%%%%%%%%

\section{Introduction}
\label{sintro}
Supernova (SN) 2011dh, discovered in the "Whirlpool Galaxy", M51, is a
recent example of a radio-loud SN.  At a distance of
$7.8^{+1.1}_{-0.9}$~Mpc \citep{Ergon2014}\footnote{\citet{Ergon2014}
  re-examined the various estimates of the distance to M51 to arrive
  at the value of $7.8^{+1.1}_{-0.9}$~Mpc, which is slightly lower
  than the value of 8.4~Mpc we had used in our earlier papers.},
SN~2011dh is also one of the nearest SNe observed in recent years.
SN~2011dh was discovered on 2011 May 31 by the amateur astronomer
Am\'ad\'ee Riou \citep{Griga+2011} \textcolor{black}{and the discovery was soon 
confirmed from pre- and post-discovery observations from various telescopes}
\citep{Griga+2011}.  The SN was coincident with the eastern spiral arm
of M51.  The explosion date is tightly constrained to be between 2011
May 31.275 and 31.893 UT \citep{Arcavi+2011b}.  We will adopt the
(rounded) midpoint of this interval, May 31.6 UT, as the explosion
date, $t_0$, and take $t$ to be the age of the SN since the explosion.

Initially, SN~2011dh was spectroscopically classified as Type IIP
\citep{SilvermanFC2011}, but further spectroscopy, which showed helium
absorption features, caused a re-classification as a Type IIb
\citep{Arcavi+2011b, Marion+2011}. A maximum expansion velocity of
$\sim20000$~\kms\ was estimated from the blue edge of the H$\alpha$
line \citep{SilvermanFC2011, Arcavi+2011b, Marion+2014}.

Radio emission was detected only a few days after the explosion, at
centimetre wavelengths \citep{Horesh+2011} with the National Radio
Astronomy Observatory\footnote{The National Radio Astronomy
  Observatory is a facility of the National Science Foundation
  operated under cooperative agreement by Associated Universities,
  Inc.} (NRAO) Karl G. Jansky Very Large Array (VLA), \textcolor{black}{at
millimetre wavelengths} \citep{HoreshZC2011} using the Combined Array
for Research in Millimeter-wave Astronomy, and at submillimetre
wavelengths \citep{Soderberg+SN2011dh-I} using the Submillimeter Array
(SMA). Results from the initial radio and
millimetre-band observations were presented in
\citet{Soderberg+SN2011dh-I}, while further broad-band measurements of
the total radio flux density as well as modelling of the lightcurve
\textcolor{black}{were} presented in \citet{Krauss+SN2011dh-II}.

A yellow supergiant which was visible in pre-explosion Hubble Space
Telescope (HST) images, but has since disappeared, has been identified
as the progenitor \citep{vDyk+2013, Ergon2014}.  The progenitor's
main-sequence mass is estimated to around 13 \Msol\ although the
various workers give masses in the range of 10 to 19
\Msol\ \citep{SahuAC2013, vDyk+2011, Maund+2011, Bersten+2012}.  It
was an extended star with a radius of $200 \sim
300$~\Rsol\ \citep{vDyk+2011, Bersten+2012, Ergon2014}.  In addition,
a blue companion to SN~2011dh's yellow supergiant progenitor has also
likely been detected in deep near-UV images obtained with the {\em
  HST} \citep{Folatelli+2014}, making SN~2011dh the second
core-collapse SN, after SN~1993J \citep{Maund+2004, Fox+2014}, to show
strong evidence of a binary companion \textcolor{black}{for} the progenitor.

The size and expansion velocity of the shock front is a basic
characteristic distinguishing different SNe, and it is therefore
important to determine it observationally as directly as
possible. VLBI observations are the most direct way of making this
measurement \citep[see e.g.,][]{Marti-Vidal+2011a,
    SN2009bb-VLBI, Brunthaler+2010a, Bietenholz2008}. 
Unlike the optical emission, which mostly originates in the denser and
more slowly moving inner ejecta, the radio emission generally traces
the fastest ejecta. The radio emission is thought to originate in the
region between the forward and reverse shocks.  \textcolor{black}
{In the particularly well-studied case of SN~1993J and based on the highly
resolved exemplary shell of radio emission seen at late times, 
\citet{Bartel+2007-93JIV} show that there is a close relationship 
between the outer boundary of the radio emission and the 
location of the forward shock at least at times $> 1$~yr, although 
\citet{Bjornsson2015} suggests that the correspondence may not be as 
good for $t < 1$~yr.}

In the particularly
well-studied case of SN~1993J, \citet{Bartel+2007-93JIV} show that
there is a close relationship between the outer boundary of the radio
emission and the location of the forward shock.

VLBI observations of SN~2011dh were first obtained at $t=14$~d
\citep[2011 Jun 14,][]{Marti-Vidal+2011d}.  Although an accurate
centre position was obtained, no useful constraint on the size could
be obtained at this early epoch
Further VLBI observations were obtained at epochs $t = 83$ and 179~d
(2011 Aug.\ 22 and 2011 Nov.\ 26) by Bietenholz et
al.\ 2012\ \citep[see
  also][]{SN2011dh_ATel}. \nocite{Bietenholz+SN2011dh-III} At both
epochs, we used the High Sensitivity Array (HSA), consisting of the
NRAO VLBA, the Effelsberg telescope and the Robert C. Byrd Green Bank
Telescope, and obtained useful measurements or upper limits on the
source size.  At $t = 83$~d we observed at 22~GHz, while at $t =
179$~d we observed at 8.4~GHz.  At $t = 179$~d, the measured outer
radius corresponds to an expansion velocity of $20000 \pm
6500$~\kms\ at 7.8~Mpc.
\citet{Rampadarath+2015} also detected SN~2011dh in
their 1.6~GHz, wide-field VLBI observations of M51 at $t = 160$~d
(2011 Nov.\ 7), but the source was unresolved, so again no useful
constraint on the source radius was obtained.

Aside from the relatively direct measurement using VLBI, the radius of
the radio emitting region can also be determined from the radio
spectral energy distribution (SED) if the spectrum is dominated by
synchrotron self-absorption (SSA), as is generally expected at early
times \citep[see][]{ChevalierF2006}. \citet{Soderberg+SN2011dh-I} and
\citet{Krauss+SN2011dh-II} show that the radio spectrum of SN~2011dh
was consistent with being dominated by SSA, and gave radius
determinations up to $t = 92$~d. Although the calculation of the
radius based on the SSA spectrum is fairly robust, it is more
model-dependent than the more direct VLBI measurements. SN~2011dh
represents so far the best example for directly comparing the radii of
the shock wave determined in these two different fashions, and
\citet{Bietenholz+SN2011dh-III} show that for the measurements up to
$t = 179$~d, there is excellent agreement between the two methods,
thereby providing important confirmation for the radii derived from
the SED by assuming SSA.

SN~2011dh was unusual in that it remained radio-bright enough for VLBI
observations for more than a year.  We undertook a further epoch of
VLBI observations, as well as observations to measure the total flux
density with the VLA to determine the continued evolution of this SN,
and we report on these results in this paper.

\section{Observations and Results} 
\subsection{VLA flux density observations and results}
\label{svla}

We observed SN~2011dh with the VLA to get a total flux density
measurements at 8.4 GHz on 2012 Aug 1 (program 12A-286) and on 2014
Jan 31 (program 13A-370).

The 2012 Aug 1 observations were done with the array in the B
configuration and we used a bandwidth of 1~GHz centred on 8.5~GHz.
They were reduced using the NRAO's Astronomical Image Processing
System (AIPS), with the flux density scale being set from observations
of 3C286 using the Perley-Butler 2010 coefficients.  We measured an
8.5-GHz flux density of $0.88 \pm 0.06$~mJy, where the uncertainty
includes the noise as well as an assumed 5\% uncertainty in the flux
density calibration.

The 2014 Jan 31 observations were done with the array in the BnA
configuration and we used a bandwidth of 1~GHz centred on 7.1 MHz.
They were reduced using AIPS, and calibrated to the same flux density
scale using observations of 3C~286. We measured a flux density of
$0.66 \pm 0.035$~mJy, where again the uncertainty includes the noise
as well as an assumed 5\% uncertainty in the flux density calibration.

\begin{figure}
\centering \includegraphics[width=
  0.46\textwidth]{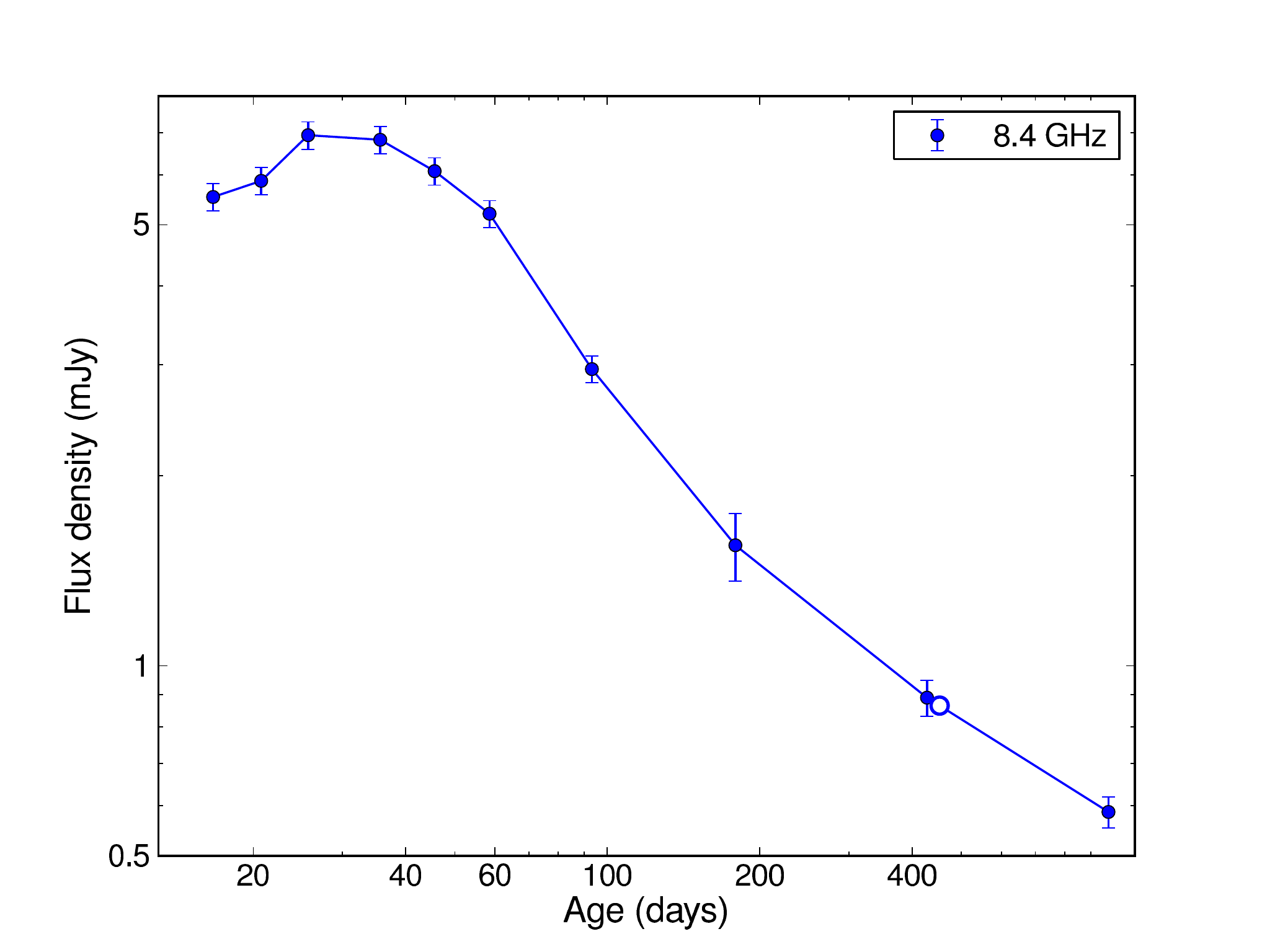}
\caption{The 8.4-GHz lightcurve of SN~2011dh, as obtained from VLA
  observations.  Flux density measurements are shown by solid circles,
  and the logarithmically interpolated value at the time of our VLBI
  observations (0.86 mJy on 2012 Aug.\ 26) by an open circle.  Details
  for the last two measurements are given in the text, and the
  remainder are taken or interpolated from
  \citet{Bietenholz+SN2011dh-III} and \citet{Krauss+SN2011dh-II}.  The
  age of the SN is calculated from the explosion date of 2011 May 31.6
  UT.}
\label{flightcurve}
\end{figure}

In Figure~\ref{flightcurve} we show the 8.4-GHz radio lightcurve of
SN~2011dh.  We include the two new measurements described above as
well as earlier ones taken from \citet{Bietenholz+SN2011dh-III} and
\citet{Krauss+SN2011dh-II}.  The logarithmically interpolated value at
the time of our VLBI observations (2012 Aug.\ 26) was 0.86~mJy.  We
scaled all the flux-density measurements taken between 7 and 9 GHz to
8.4 GHz using a spectral index of $\alpha -0.7$ \citep[where $S_\nu
  \propto \nu^\alpha$][]{Krauss+SN2011dh-II}.  We are not sensitive to
the exact value assumed for the spectral index: the earlier
measurements, when the spectral index may have been notably different
from $-0.7$ were taken at 8.4 GHz, and therefore required no scaling,
and even in the most extreme case of the last
measurement, at 7.1~GHz, a difference of 0.2 in the spectral index
would change the plotted value by less than the uncertainty.
A logarithmic fit to the values after $t = 45$~d gives
an average flux density decay rate, $\beta$, of $-0.79 \pm 0.05$,
where $S \propto t^{-\beta}$.  The lightcurve appears to flatten
after $t = 179$~d, with a two-segment fit giving $\beta = 1.17 \pm
0.13$ for $t < 179$~d to $\beta = 0.57 \pm 0.13$ for $t > 179$~d.

\subsection{VLBI Observations}
\label{sobs-vlbi}
We obtained VLBI imaging observations of SN~2011dh using the
high-sensitivity array, which consisted of the NRAO Very Long Baseline
Array (VLBA; $10 \times 25$-m diameter, distributed across the United
States) and the NRAO Robert C. Byrd ($\sim$105~m diameter) telescope
and the Effelsberg (100~m diameter) telescope.  Due to a technical
failure, the St.\ Croix VLBA telescope did not observe.  The
observations occurred on 2012 Aug 26, and lasted for a total of 13~hr.
At the midpoint of our observing session the age of the SN was 453~d.

To measure the tropospheric zenith delay and clock offsets at each
antenna we included three ``geodetic blocks'' of $\sim$30~minutes each
at the start, middle and end of our observations
\citep[see][]{BrunthalerRF2005,ReidB2004}.  In each of these geodetic
blocks we observed 9-14 bright reference sources chosen from the
International Celestial Reference Frame (ICRF) list of sources
\citep{FeyGJ2009}. For our geodetic blocks we recorded 8 intermediate
frequencies covering a bandwidth of 8~MHz spread over a $\sim$400~MHz
range.

For the observations of SN~2011dh we recorded a contiguous bandwidth
of 64~MHz in each of the two senses of circular polarization with
two-bit sampling for a total bit rate of 512~Mbit~s$^{-1}$.  We used
the same calibrator sources as in our previous VLBI observations of
SN~2011dh. We used J1332+4722 (ICRF J133245.2+472222), which is
0.5\arcdeg\ away from SN~2011dh, as our primary calibrator.  This
source is an ICRF source with a position known to
$\sim$70~\muas\ \citep{FeyGJ2009}.  Any positions in this paper are
calculated by taking the position of J1332+4722 to be Right Ascension
(RA) = \RA{13}{32}{45}{24642}, Declination (decl.) =
\dec{47}{22}{22}{6670}. We also obtained observations of the quasar,
JVAS J1324+4743, about 1.5\arcdeg\ away from J1332+4722, as an
astrometric check source. The purpose of the observations of
J1324+4743 was to check the quality of the phase-referencing and also
to provide a second astrometric reference source to check the
positional accuracy.

We used a cycle time of $\sim$150~s, with $\sim$110~s on SN~2011dh and
$\sim$40~s on J1332+4722. During the observing run we spent three
$\sim$20~m periods observing our astrometric check source, J1324+4743,
using a similar phase-referencing pattern as we used for SN~2011dh.

The VLBI data were correlated with the DiFX correlator
\citep{Deller+2011}, and the analysis carried out
with AIPS.  We corrected for the dispersive ionospheric delay using
the AIPS task TECOR, and we solved for the zenith tropospheric delay
on the basis of our geodetic-block observations. We discarded any
SN~2011dh visibility data obtained when either of two of the antennas
involved was observing at elevations below 10\arcdeg.

The initial flux density calibration was done through measurements of
the system temperature at each telescope, and then improved through
self-calibration of the primary reference source J1332+4722.  This
source is slightly resolved, as can be seen on the images in the VLBA
calibrator list
data-base\footnote{\url{http://www.vlba.nrao.edu/astro/calib}}, and
also from our previous observations at 22 and 8.4~GHz
\citep{Bietenholz+SN2011dh-III}. We see a similar structure in the
image from our current observations at 8.4~GHz, where a weak extension
or second component is visible $\sim$2~mas to the west-southwest of
the peak.

Our final amplitude and phase calibration was derived using a CLEAN
model of this source, with the peak-brightness point in the image
being placed at the nominal coordinates given above\footnote{The
  deviation of the source geometry of J1332+4722 from a point source
  is small enough so that the effect of using a point model in the
  solutions for delay and delay rate made using FRING is negligible.}.
Finally this calibration was interpolated to the intervening scans of
SN~2011dh.

\subsection{VLBI Results}
\label{sresults-VLBI}
\begin{figure}
\centering \includegraphics[width=
  0.46\textwidth]{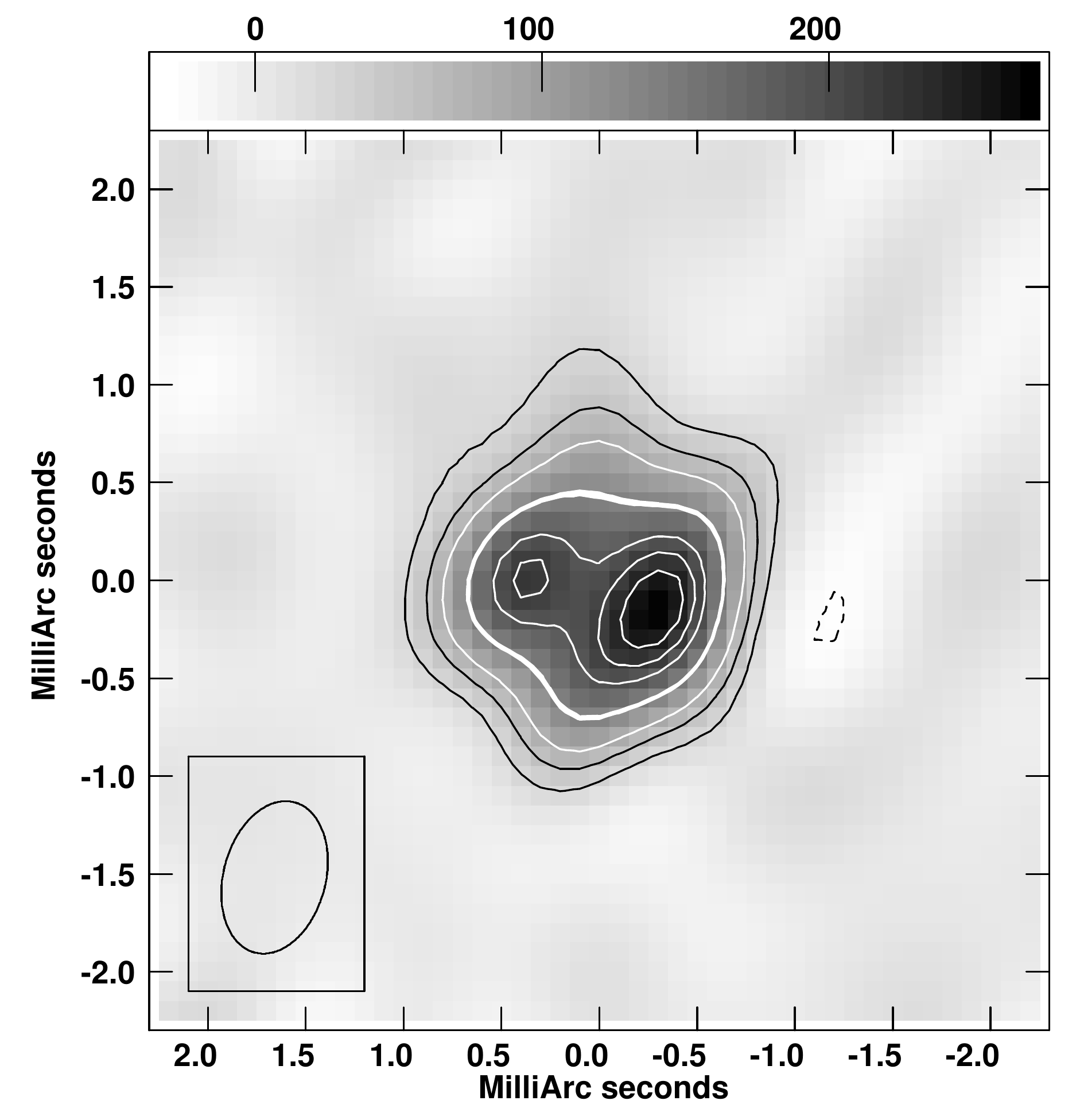}
\caption{The 8.4-GHz VLBI CLEAN image of SN~2011dh on 2012 August 26,
  at $t = 453$~d. The greyscale and contours both show
    the brightness.  The greyscale is labelled in $\mu$\Jb, and the
    contours are drawn at $-$10, 10, 20, 30, 50 (emphasized), 70, 80
    and 90\% of the peak brightness, which was 270~$\mu$\Jb. The
    convolving beam, indicated at lower left, was $0.79 \times
    0.52$~mas (FWHM) at p.a.\ $-15$\arcdeg. The image was made using
    Briggs' robust weighting \citep{BriggsSS1999}, with the weighting
    set to approximately midway between uniform and natural weighting
    (AIPS robustness parameter set to $-1$). The total CLEAN flux
    density was 760~$\mu$Jy, and the background rms level was
    12~$\mu$\Jb. North is up and east is to the left, and the
    coordinates are given as offsets from the centre position of the
    fit model, which is at RA = \RA{13}{30}{5}{1055478} and decl.\ =
    \dec{47}{10}{10}{92261} (see Text \ref{sresults-VLBI}).}
\label{figure1}
\end{figure}

We detected SN~2011dh with a sufficient
signal-to-noise ratio to obtain a high-quality VLBI image with a
resolution of $0.79 \times 0.52$~mas (FWHM), a dynamic range of 23,
and an image background rms of 12~\muJb. We show the image in Figure
\ref{figure1}.  The total CLEAN flux density was 760~$\mu$Jy, which
was $\sim$90\% of the total flux density measured using the VLA (see
section \ref{svla}). The structure of SN~2011dh is relatively circular
at the lower contours, but two hot-spots, bilaterally
located approximately east and west, are evident in the image.

As in our previous VLBI observations \citep{SN2011dh_ATel,
  Bietenholz+SN2011dh-III}, we again turn to fitting a spherical-shell
model, with an outer radius of $1.25 \times$ the inner radius,
directly to the visibility measurements to accurately measure the size
and centre position of the source.  The free parameters in the fit are
the centre position, the radius, and the flux density.

We take the fitted centre position of the model as our best estimate
of the centre position of SN~2011dh.  For the observations of
SN~2011dh, at $t = 453$~d, the position of the model centre was RA =
\RA{13}{30}{5}{1055478} and decl.\ = \dec{47}{10}{10}{92261}. The
statistical uncertainty on this position is $30~\muas$ in both RA and
decl.

The fitted outer radius was $636 \pm 29$~\muas. The statistical
uncertainty was $\pm 23$~\muas, and a Monte-Carlo simulation with 16
trials where we randomly varied the amplitude gains by $\sigma = 20$\%
gave an additional radius uncertainty due to possible amplitude
calibration errors of 13~\muas. We add these contributions in
quadrature, and our final standard error on the fitted outer angular
radius is therefore 29 \muas.  At $7.8^{+1.1}_{-0.9}$~Mpc, this angular 
radius corresponds to a linear radius of $(7.4^{+1.1}_{-0.9}) \times 10^{16}$~cm, 
and an average expansion velocity since the explosion of
$19000^{+2800}_{-2400}$~\kms, where our uncertainties on the linear
radius and expansion velocity includes the contribution of the
uncertainty of the distance.

\section{Discussion}
\label{sdiscuss}
We obtained phased-referenced VLBI observations of SN~2011dh, with the
primary goal of obtaining a clearly resolved image of the expanding
shell of ejecta, at $t=453$~d after the explosion. We obtained a
fairly well resolved image of SN~2011dh, shown in Fig~\ref{figure1},
which shows a shell like structure that is fairly circular.  Our image
shows a bilateral enhancement of brightness, with hot-spots located
approximately east and west along the ridge.

In the case of the similar Type IIb SN, SN~1993J, there are clear
time-dependent modulations of the brightness along the ridge-line
which were larger than the noise or any systematic
effect during the first two years \citep{SN93J-3,
Bietenholz2008}.  Do the hot-spots in the image of SN~2011dh
indicate a real brightness enhancement along the ridge?

Even in the case of a completely circular structure, such a bilateral
enhancement is in fact expected in the case of an elliptical CLEAN
beam, with the hot-spots occurring at a p.a.\ at right angles to the
elongation of the beam.  We made simulated visibility data sets of a
purely circular shell model with a noise level similar to that in the
real data.  When these were CLEANed in a fashion similar to the real
data, the resulting images also tended to display
apparent hot spots, similar in brightness to those
visible in Figure~\ref{figure1}.  We therefore conclude that although
the bilateral enhancement seen in Figure~\ref{figure1} could be real,
a perfectly uniformly circular structure is also compatible with the
data. An enhancement on the west side is suggested,
but not demanded by the data.

Optical polarization measurements of SN~2011dh suggest
significant departures from sphericity for times $t < 30$~d
\citep{Mauerhan+2015}, which is in contrast to the relatively
circular outline seen in our VLBI image. A similar contrast between
a circularly symmetric VLBI image and optical polarization
suggesting significant departures from symmetry exists in SN~1993J.
The reason for the contrast is likely that the radio emission arises
from the outer shock, which is relatively spherical, while the
optical polarization arises from the inner ejecta, which are much
less so.

We combined the present VLBI measurement of the centre position and
outer radius with those from earlier observations from
\citet{Bietenholz+SN2011dh-III}, \citet{Krauss+SN2011dh-II} and
\citet{Marti-Vidal+2011d}.

Since slightly different correlator positions for the phase-reference
source J1322+4722 were used in the different VLBI experiments, we
corrected the phase-referenced SN~2011dh centre position as well as
that of the check source, J1324+4743, so that they are all relative to
the position of J1322+4722 given in \S \ref{sobs-vlbi} above.

Since all three of our VLBI runs included observations of the check
source, we can compare the phase-referenced position of J1324+4743
among the three epochs.  We found that the scatter in this position
over the three epochs was 130 \muas\ in R.A.  and 90 \muas\ in decl.
This scatter is larger than the expected astrometric errors,
suggesting that there is some small apparent motion in one (or both)
of the sources.  Since we do not know whether our phase-reference
source, J1324+4743, or the check source J1322+4722 has an apparently
variable position, we take this scatter to be the uncertainty in the
position of the phase-reference source.  Based on this, we then take a
conservative value of 130~\muas\ for the uncertainty also in the
centre position of SN~2011dh.

Turning now to the centre positions of SN~2011dh, and including the
value at $t=14$~d from \citet{Marti-Vidal+2011d}, we performed a
least-squares fit of the position as a function of time. We find a
proper motion of $55 \pm 66$~\muasyr\ (at p.a.\ $-15$\arcdeg), which
corresponds to $2000 \pm 2400$~\kms, with a $3\sigma$ upper limit of
9200~\kms.

We note that any proper motion due M51's galactic rotation is expected
to be only a few hundred \kms, well below our uncertainty, so we can
take this projected speed to be that of SN~2011dh's centre with
respect to its local frame of rest within M51.

\begin{figure}
\centering \includegraphics[width=
  0.5\textwidth]{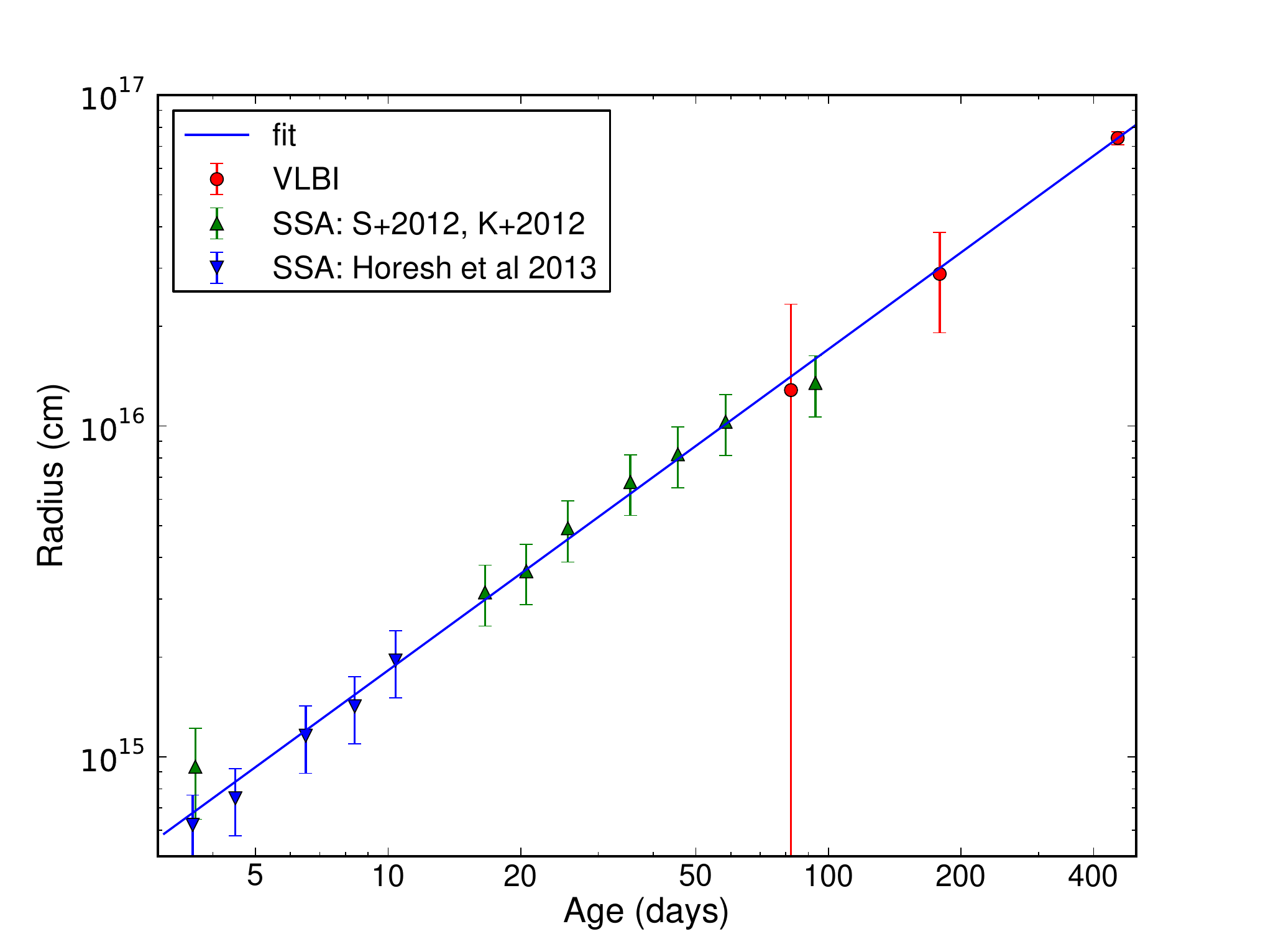}
\caption{The shock front radii of SN~2011dh as estimated by two
  independent methods.  Using \textcolor{black}{red circles}, we plot the values derived
  from fitting spherical shell models to the VLBI visibility-data from
  this paper and \citet{Bietenholz+SN2011dh-III}.  The plotted
  $1\sigma$ error bars include statistical and systematic
  contributions (see Section~\ref{sresults-VLBI} for details) but do
  not include the contribution from the distance, since any error in
  the distance would affect all the points in the same way.  It should
  be noted that the error bars on the last point are smaller than the
  point symbol and is therefore not visible in the
  plot. Using \textcolor{black}{triangles}, we plot the values calculated
    from the radio spectral energy distribution under the assumption
    that it is dominated by SSA (synchrotron self-absorption), with
    \textcolor{black}{green triangles pointing up} for the values taken from
    \citet[][``S+2012'']{Soderberg+SN2011dh-I} \textcolor{black}{and} 
    \citet[][``K+2012'']{Krauss+SN2011dh-II} and \textcolor{black}{blue triangles pointing down}
     for those from \citet{Horesh+2013}.  The uncertainties include
    statistical and systematic components and all radius measurements
    have been rescaled to our distance of 7.8~Mpc.  The blue line
  represents our power-law fit to all the values, with
 $R = 5.30\times10^{15}\;(t/{\rm 30\, d})^{0.97}$~cm
    (see text).}

\label{fexpansioncurve}
\end{figure}

For the outer radius of the radio emitting region, which as we have
argued, should be closely tied to the outer shock radius, we use the
value determined in the present work along with the two VLBI values
from \citet{Bietenholz+SN2011dh-III}, and those determined from the
radio spectrum by assuming SSA from \citet{Krauss+SN2011dh-II}, as
as well as from \citet{Horesh+2013} ($t \sim 4 to11$~d). 
We scale all the radii to a consistent distance of
7.8~Mpc.  We plot them in Figure \ref{fexpansioncurve}.

The expansion of a SN is usually parametrized as a power-law with $R =
At^m$, where $R$ is the radius, $t$ is the time, $m$ is the power-law
index, called the ``deceleration parameter'' and $A$ is the radius at
$t = 1$.  A weighted least-squares fit to all the values of $R$ gives
$$R = (5.3 \pm 0.1) \; (t/30 \; {\mathrm d})^{(0.97
\pm 0.01)} \; \times 10^{15} \; \mathrm{cm}$$

The uncertainty on $m$ is the statistical one only,
 calculated assuming that the points are all independent. All the
radius measurements are consistent with a power-law evolution.  The
fitted expansion index $m=0.97 \pm 0.01$ implies
almost free expansion.

As already noted in \citet{Bietenholz+SN2011dh-III}, the radii
measured with VLBI are consistent with those determined somewhat less
directly from the SED under the assumption of SSA.
We further note that the SSA radii derived by
\citet{Horesh+2013} are quite consistent with those of
\citet{Soderberg+SN2011dh-I} and \citet{Krauss+SN2011dh-II}.
The earlier VLBI radius measurements were rather more uncertain than
those obtained from the SED.  Our new measurement, by contrast, is
rather more precise, and is still consistent with the
powerlaw implied by the SED-based radius measurements.
Note that at $t > 100$~d, the turnover frequency
becomes too low to be directly measured, thus it is not possible to
obtain radii from SED fits at the times of our more accurate VLBI
measurements.  Nonetheless,
while it is possible that the SED-derived radii are systematically too
small and that the expansion decelerated after $t \sim 100$~d, we
consider it an unlikely coincidence that the measured VLBI radius at
$t = 453$~d would lie so close to the extrapolation of the powerlaw
seen at early times in that case.

The radii determined from the SED depend on $D^{(18/19)}$ whereas the
(linear) radii determined from the VLBI measurements
depend on $D$, so only very large errors in $D$ would produce a
substantial scaling between the two kinds of radius determinations.
We note, however, that by comparing the angular expansion velocity as
determined from VLBI and less directly from the SED fits, with linear
expansion velocity measured from optical spectroscopy, a direct
distance can be determined via the ``expanding shock front'' method.
See \citet{Bartel+2007-93JIV} for a description of
this method, and its application in the case of SN~1993J in M81\@.  We
defer a detailed comparison along these lines to a future paper, but
note here that our fit implies an expansion velocity at $t = 30$~d of
$20000^{+2900}_{-2400}$~\kms, where again our
uncertainty on the expansion velocity includes the contribution from
the distance uncertainty (+15\%, $-12$\%).
This velocity compares very well to maximum velocity seen at the blue
edge of H$\alpha$ absorption seen in the optical at the same age
\citep[e.g.][]{Marion+2014}.

\subsection{Self-similar approximation and estimate of CSM and ejecta density profiles}
\label{sself-similar}

The amount of radio emission as well as the
deceleration are ultimately dependent on the energy dissipated as
the shock ploughs out through the CSM, and depend therefore on the
density profiles in the ejecta and the CSM. In particular, if the
density profiles of the ejecta and the CSM are power-laws in radius,
then a self-similar solution exists, and both the total flux-density
and the shock radius follow powerlaws in time such that the flux
density decays as $S \propto t^\beta$ while $r \propto t^m$
\citep{Chevalier1982b, FranssonLC1996}.

Our fit to the expansion of SN~2011dh gives $m = 0.97
\pm 0.01$, and our fit to the optically thin part of the 8.4-GHz
lightcurve gives $\beta = -0.79 \pm 0.05$.  If we assume power-law
density profiles, with the ejecta density being $\propto r^n$ and
the CSM density $\propto r^s$, and taking the radio spectral index
to be $\alpha = -0.7 \pm 0.2$, then those values of $m$ and $\beta$
imply $n = 31 \pm 11$ and $s = 2.0 \pm 0.1$.  So the implied CSM
density profile overall is consistent with a wind density profile
with $\rho \propto r^{-2}$.  The ejecta density profile is less well
determined, but the small deceleration and flat lightcurve imply a
very steep profile in the ejecta, consistent with what was found by
\citet{Maeda2012} based on X-ray observations.

Although our radius measurements are compatible with
an unbroken powerlaw, the measurements suggest a flattening after $t
= 179$~d instead of a single powerlaw, with $\beta = 1.17 \pm 0.13$
until then and $0.57 \pm 0.13$ thereafter.  No obvious corresponding
change is seen in the expansion curve.  A flattening of the flux
density decay implies either a steeper density profile in the
ejecta, or a flatter one in the CSM\@.  The former would be
accompanied by \textcolor{black}{a} decrease in deceleration while the latter would be
accompanied by an increase.

Since we see no obvious change in the deceleration at
$t = 179$~d, our measurements do not reliably distinguish between
these two cases.  Although the average deceleration over all the
measurements is well determined at $m = 0.97 \pm 0.01$, the value
for $t > 179$~d is much less so, and a change of up to 0.1 in $m$ at
$t > 179$~d is compatible with our measurements.  Since $m > 1$ can
only happen with some form of continued energy input, we think it more
likely that $m$ has decreased at the latest times.

We note that a density structure more complicated than
a simple powerlaw is not unreasonable for either the ejecta nor the
CSM, and that in fact departures from powerlaw density structure 
\textcolor{black}{have} been suggested for both the ejecta and the CSM of the well-studied
Type IIb SN~1993J.  We will elaborate on the comparison between
SN~2011dh and SN~1993J in the following section (\S
\ref{scompare93J}).

\subsection{Comparing SN~2011dh to SN~1993J.}
\label{scompare93J}

We can compare SN~2011dh to SN~1993J: both were of Type IIb, and had
extended progenitors with binary companions, and had lost a
considerable fraction of their hydrogen envelopes to the companion
before the SN explosion.  For SN~1993J, the radio lightcurves and
expansion were particularly well determined \citep[see,
  e.g.][]{Bartel+2002-93JII, Marti-Vidal+2011b}.  SN~1993J was about
four times as radio luminous as SN~2011dh, as the former reached a
peak 8.4-GHz spectral luminosity of $\sim 2\times 10^{27}$~\ergsHz,
compared to only $0.5 \times 10^{27}$~\ergsHz\ for the latter.  The
evolution of SN~1993J's radio lightcurves was much slower than that of
SN~2011dh, since the former reached an 8.4~GHz peak only after $t \sim
150$~d, compared to the $t = 30$~d for SN~2011dh.  However, we note
that a circumstellar density profile flatter with $s < 2$,
i.e.\ flatter than that of a steady wind, was also suggested for
SN~1993J \citep[e.g.][]{vDyk+1994, FranssonLC1996}, although models
with a constant $s = 2$ can also reproduce the measurements
\citep[e.g.][]{FranssonB1998, Marti-Vidal+2011b}.

In the case of SN~1993J, departures from simple
powerlaw profiles have been suggested for both the ejecta and the
CSM.  \citet{Iwamoto+1997} modelled the optical emission, while
\citet{SuzukiN1995} modelled the early X-ray emission of SN~1993J,
and both authors found that to reproduce the observed emission
required an ejecta structure considerably more complicated than a
simple powerlaw, including an increase in density going inward past
the H/He interface.  Modelling of the radio lightcurves and
expansion by \citet{MioduszewskiDB2001} showed that such local
increases in the ejecta density can cause temporary flattening of
the lightcurve decay.  \citet{Bartel+2002-93JII} looked at the radio
lightcurve and expansion rate of SN~1993J till $t = 3164$~d and
suggest changes in the CSM density profile.

Turning to SN~2011dh, overall the fast rise of the
radio lightcurve to its maximum, the small deceleration and the low
peak luminosity imply that the CSM in this case was less dense than
that of SN~1993J.  Other Type IIb SNe which have shown similarly quick
rise times and relatively low radio luminosities are SN~2008bo
\citep{Stockdale+2008}, and SN~2008ax \citep{Roming+2009}, and indeed
the similarities of SN~2011dh's optical spectra to those of SN~2008bo
have already been pointed out \citep{Maund+2011, Marion+2014}

The small deceleration suggests that the swept-up mass
must still be low compared to that of the ejecta.
\citet{Krauss+SN2011dh-II}, \citet{Horesh+2013}, and
\citet{Maeda+2015} have all estimated mass-loss rates for
SN~2011dh's progenitor, with the largest of those three estimates
being that of Maeda et al, which was $\sim 3 \times
10^{-5}$~\Msol~yr$^{-1}$.  Using that mass-loss rate, and a wind
speed for the yellow supergiant of 20~\kms, we calculate that
the swept-up mass at $t = 453$~d was $\sim 1.3 \times 10^{-3}
\Msol$. If we used the mass-loss rates of \citet{Krauss+SN2011dh-II}
or \citet{Horesh+2013} we would derive a somewhat lower 
swept-up mass.  Since the ejected mass is still low with the mass of
the ejecta (a few $\Msol$) at $t = 453$~d, our finding that the SN
is still in almost free expansion is consistent with the
expectations.

\section{Conclusions}
\label{sconclud}

\begin{trivlist}

\item{1.} We obtained a new, resolved, VLBI image of
  SN~2011dh, at $t = 453$~d after the explosion.  We found that
  SN~2011dh shows a relatively circular projected shell structure,
  possibly with some \textcolor{black}{enhancement on} the west side.  With our new
  VLBI image, SN~2011dh takes its place among the only six recent SNe
  for which resolved images of the ejecta are available.  With our new
  VLBI image, SN~2011dh takes its place among the only six recent SNe
  for which resolved images of the ejecta are available --- the others
  being SN 1979C \citep{SN79C-shell}, SN~1987A \citep{Ng+2011},
  SN~1986J \citep{SN86J-2}, SN~1993J \citep{SN93J-3,
      Marcaide+1997}, SN~2008iz \citep{Brunthaler+2010a}. 

\item{2.} We also obtained measurements of the total flux density with
  the VLA, and the radio lightcurve shows an approximately power-law
  decay with $S \propto t^{-0.79}$), \textcolor{black}{although the} decay flattens
  somewhat after $t = 179$~d.

\item{3.} Our astrometric VLBI measurements \textcolor{black}{give} an
  upper limit of 9200~\kms\ on the average projected speed of the
  centrepoint of SN~2011dh since the explosion.

\item{4.}  \textcolor{black}{We measured a radius of
  SN~2011dh at $t = 453$~d of $(7.4^{+1.1}_{-0.9}) \times 10^{16}$~cm
  using our VLBI data.}
  Comparing to earlier values derived both from VLBI observations and
  from fitting of the SED, we show that SN~2011dh's expansion is well
  fit with a power-law with \textcolor{black}{$R \propto t^{0.97\pm0.10}$.}  
  The SN is therefore still in almost free expansion.

\item{5.} Our fit to the expansion implies a velocity
  for the forward shock at $t = 30$~d of $20000^{+2900}_{-2400}$~\kms,
  which is comparable to the highest velocities seen in optical
  spectral lines.

\item{6.} The expansion curve and the radio lightcurve
  together are consistent with a wind density profile \textcolor{black}{($\rho \propto
  r^{-2}$)} in the circumstellar medium, but suggest a very steep
  density profile in the ejecta.

\end{trivlist}

%%%%%%%%%%%%%%%%%%%%%%%%%%%%%%%%%%%%%%%%%%%%%%%%
%  define journalnames abbreviations, req. for bibtex.

\newcommand{\araa}{Ann.\ Rev.\ Astron.\ Astrophys.}
\newcommand{\aap}{Astron.\ Astrophys.}
\newcommand{\aapr}{Astron.\ Astrophys.\ Rev.}
\newcommand{\aaps}{Astron.\ Astrophys.\ Suppl.\ Ser.}
\newcommand{\aj}{AJ} \newcommand{\apj}{ApJ} \newcommand{\apjl}{ApJL}
\newcommand{\apjs}{ApJS} \newcommand{\apss}{ApSS}
\newcommand{\baas}{BAAS} \newcommand{\memras}{Mem.\ R. Astron.\ Soc.}
\newcommand{\memsai}{Mem.\ Soc.\ Astron.\ Ital.}
\newcommand{\mnras}{MNRAS} \newcommand{\iaucirc}{IAU Circ.}
\newcommand{\jrasc}{J.\ R.\ Astron.\ Soc.\ Can.}
\newcommand{\nat}{Nat} \newcommand{\pasa}{PASA}
\newcommand{\pasj}{PASJ} \newcommand{\pasp}{PASP}

\bibliographystyle{mn2e} 
\bibliography{sn2011bib}

\begin{thebibliography}{54}
\expandafter\ifx\csname natexlab\endcsname\relax\def\natexlab#1{#1}\fi

\bibitem[{{Arcavi} {et~al}\mbox{.}(2011){Arcavi}, {Gal-Yam}, {Yaron},
  {Sternberg}, {Rabinak}, {Waxman}, {Kasliwal}, {Quimby}, {Ofek}, {Horesh},
  {Kulkarni}, {Filippenko}, {Silverman}, {Cenko}, {Li}, {Bloom}, {Sullivan},
  {Nugent}, {Poznanski}, {Gorbikov}, {Fulton}, {Howell}, {Bersier}, {Riou},
  {Lamotte-Bailey}, {Griga}, {Cohen}, {Hachinger}, {Polishook}, {Xu},
  {Ben-Ami}, {Manulis}, {Walker}, {Maguire}, {Pan}, {Matheson}, {Mazzali},
  {Pian}, {Fox}, {Gehrels}, {Law}, {James}, {Marchant}, {Smith}, {Mottram},
  {Barnsley}, {Kandrashoff}, \& {Clubb}}]{Arcavi+2011b}
{Arcavi} I. {et~al.}, 2011, \apjl, 742, L18

\bibitem[{{Bartel} \& {Bietenholz}(2008)}]{SN79C-shell}
{Bartel} N., {Bietenholz} M.~F., 2008, \apj, 682, 1065

\bibitem[{{Bartel} {et~al}\mbox{.}(2002){Bartel}, {Bietenholz}, {Rupen},
  {Beasley}, {Graham}, {Altunin}, {Venturi}, {Umana}, {Cannon}, \&
  {Conway}}]{Bartel+2002-93JII}
{Bartel} N. {et~al.}, 2002, \apj, 581, 404

\bibitem[{{Bartel} {et~al}\mbox{.}(2007){Bartel}, {Bietenholz}, {Rupen}, \&
  {Dwarkadas}}]{Bartel+2007-93JIV}
{Bartel} N., {Bietenholz} M.~F., {Rupen} M.~P., {Dwarkadas} V.~V., 2007, \apj,
  668, 924

\bibitem[{{Bersten} {et~al}\mbox{.}(2012){Bersten}, {Benvenuto}, {Nomoto},
  {Ergon}, {Folatelli}, {Sollerman}, {Benetti}, {Botticella}, {Fraser},
  {Kotak}, {Maeda}, {Ochner}, \& {Tomasella}}]{Bersten+2012}
{Bersten} M.~C. {et~al.}, 2012, \apj, 757, 31

\bibitem[{{Bietenholz}(2008)}]{Bietenholz2008}
{Bietenholz} M., 2008, in Proc. 9th European VLBI Network Symposium and EVN
  Users Meeting: The role of VLBI in the Golden Age for Radio Astronomy. PoS,
  Bologna, Italy, available at:
  http://pos.sissa.it//archive/conferences/072/064/IX$\%20\\
  $EVN$\%20$Symposium\_064.pdf

\bibitem[{{Bietenholz}, {Bartel} \& {Rupen}(2003){Bietenholz}, {Bartel}, \&
  {Rupen}}]{SN93J-3}
{Bietenholz} M.~F., {Bartel} N., {Rupen} M.~P., 2003, \apj, 597, 374

\bibitem[{{Bietenholz}, {Bartel} \& {Rupen}(2010){Bietenholz}, {Bartel}, \&
  {Rupen}}]{SN86J-2}
{Bietenholz} M.~F., {Bartel} N., {Rupen} M.~P., 2010, \apj, 712, 1057

\bibitem[{{Bietenholz} {et~al}\mbox{.}(2011){Bietenholz}, {Brunthaler},
  {Bartel}, {Chomiuk}, {Rupen}, {Soderberg}, \& {Zauderer}}]{SN2011dh_ATel}
{Bietenholz} M.~F., {Brunthaler} A., {Bartel} N., {Chomiuk} L., {Rupen} M.~P.,
  {Soderberg} A., {Zauderer} B., 2011, The Astronomer's Telegram, 3641, 1

\bibitem[{{Bietenholz} {et~al}\mbox{.}(2012){Bietenholz}, {Brunthaler},
  {Soderberg}, {Krauss}, {Zauderer}, {Bartel}, {Chomiuk}, \&
  {Rupen}}]{Bietenholz+SN2011dh-III}
{Bietenholz} M.~F., {Brunthaler} A., {Soderberg} A.~M., {Krauss} M., {Zauderer}
  B., {Bartel} N., {Chomiuk} L., {Rupen} M.~P., 2012, \apj, 751, 125

\bibitem[{{Bietenholz} {et~al}\mbox{.}(2010){Bietenholz}, {Soderberg},
  {Bartel}, {Ellingsen}, {Horiuchi}, {Phillips}, {Tzioumis}, {Wieringa}, \&
  {Chugai}}]{SN2009bb-VLBI}
{Bietenholz} M.~F. {et~al.}, 2010, \apj, 725, 4

\bibitem[{{Bjornsson}(2015)}]{Bjornsson2015}
{Bjornsson} C.~I., 2015, ArXiv: 1509.04533

\bibitem[{{Briggs}, {Schwab} \& {Sramek}(1999){Briggs}, {Schwab}, \&
  {Sramek}}]{BriggsSS1999}
{Briggs} D.~S., {Schwab} F.~R., {Sramek} R.~A., 1999, in Astronomical Society
  of the Pacific Conference Series, Vol. 180, Synthesis Imaging in Radio
  Astronomy II, {Taylor} G.~B., {Carilli} C.~L., {Perley} R.~A., eds., San
  Francisco, CA: ASP, p. 127

\bibitem[{{Brunthaler} {et~al}\mbox{.}(2010){Brunthaler}, {Mart{\'{\i}}-Vidal},
  {Menten}, {Reid}, {Henkel}, {Bower}, {Falcke}, {Feng}, {Kaaret}, {Butler},
  {Morgan}, \& {Wei{\ss}}}]{Brunthaler+2010a}
{Brunthaler} A. {et~al.}, 2010, \aap, 516, A27

\bibitem[{{Brunthaler}, {Reid} \& {Falcke}(2005){Brunthaler}, {Reid}, \&
  {Falcke}}]{BrunthalerRF2005}
{Brunthaler} A., {Reid} M.~J., {Falcke} H., 2005, in Astronomical Society of
  the Pacific Conference Series, Vol. 340, Future Directions in High Resolution
  Astronomy, {J.~Romney \& M.~Reid}, ed., p. 455

\bibitem[{{Chevalier}(1982)}]{Chevalier1982b}
{Chevalier} R.~A., 1982, \apj, 259, 302

\bibitem[{{Chevalier} \& {Fransson}(2006)}]{ChevalierF2006}
{Chevalier} R.~A., {Fransson} C., 2006, \apj, 651, 381

\bibitem[{{Deller} {et~al}\mbox{.}(2011){Deller}, {Brisken}, {Phillips},
  {Morgan}, {Alef}, {Cappallo}, {Middelberg}, {Romney}, {Rottmann}, {Tingay},
  \& {Wayth}}]{Deller+2011}
{Deller} A.~T. {et~al.}, 2011, \pasp, 123, 275

\bibitem[{{Ergon} {et~al}\mbox{.}(2014){Ergon}, {Sollerman}, {Fraser},
  {Pastorello}, {Taubenberger}, {Elias-Rosa}, {Bersten}, {Jerkstrand},
  {Benetti}, {Botticella}, {Fransson}, {Harutyunyan}, {Kotak}, {Smartt},
  {Valenti}, {Bufano}, {Cappellaro}, {Fiaschi}, {Howell}, {Kankare}, {Magill},
  {Mattila}, {Maund}, {Naves}, {Ochner}, {Ruiz}, {Smith}, {Tomasella}, \&
  {Turatto}}]{Ergon2014}
{Ergon} M. {et~al.}, 2014, \aap, 562, A17

\bibitem[{{Fey}, {Gordon} \& {Jacobs}(2009){Fey}, {Gordon}, \&
  {Jacobs}}]{FeyGJ2009}
{Fey} A.~L., {Gordon} D., {Jacobs} C.~S., eds., 2009, IERS Technical Note,
  Vol.~35, {The Second Realization of the International Celestial Reference
  Frame by Very Long Baseline Interferometry}. Frankfurt: Verlag des Bundesamts
  f\"ur Kartographie und Geod\"asie, p.~1

\bibitem[{{Folatelli} {et~al}\mbox{.}(2014){Folatelli}, {Bersten}, {Benvenuto},
  {Van Dyk}, {Kuncarayakti}, {Maeda}, {Nozawa}, {Nomoto}, {Hamuy}, \&
  {Quimby}}]{Folatelli+2014}
{Folatelli} G. {et~al.}, 2014, \apjl, 793, L22

\bibitem[{{Fox} {et~al}\mbox{.}(2014){Fox}, {Azalee Bostroem}, {Van Dyk},
  {Filippenko}, {Fransson}, {Matheson}, {Cenko}, {Chandra}, {Dwarkadas}, {Li},
  {Parker}, \& {Smith}}]{Fox+2014}
{Fox} O.~D. {et~al.}, 2014, \apj, 790, 17

\bibitem[{{Fransson} \& {Bj{\"o}rnsson}(1998)}]{FranssonB1998}
{Fransson} C., {Bj{\"o}rnsson} C.-I., 1998, \apj, 509, 861

\bibitem[{{Fransson}, {Lundqvist} \& {Chevalier}(1996){Fransson}, {Lundqvist},
  \& {Chevalier}}]{FranssonLC1996}
{Fransson} C., {Lundqvist} P., {Chevalier} R.~A., 1996, \apj, 461, 993

\bibitem[{{Griga} {et~al}\mbox{.}(2011){Griga}, {Marulla}, {Grenier}, {Sun},
  {Gao}, {Lamotte Bailey}, {Koff}, {Mikuz}, {Dintinjana}, {Silverman}, {Cenko},
  {Filippenko}, {Li}, {Yamanaka}, {Itoh}, {Arai}, {Nagashima}, \&
  {Kajiawa}}]{Griga+2011}
{Griga} T. {et~al.}, 2011, Central Bureau Electronic Telegrams, 2736, 1

\bibitem[{{Horesh} {et~al}\mbox{.}(2013){Horesh}, {Stockdale}, {Fox}, {Frail},
  {Carpenter}, {Kulkarni}, {Ofek}, {Gal-Yam}, {Kasliwal}, {Arcavi}, {Quimby},
  {Cenko}, {Nugent}, {Bloom}, {Law}, {Poznanski}, {Gorbikov}, {Polishook},
  {Yaron}, {Ryder}, {Weiler}, {Bauer}, {Van Dyk}, {Immler}, {Panagia},
  {Pooley}, \& {Kassim}}]{Horesh+2013}
{Horesh} A. {et~al.}, 2013, \mnras, 436, 1258

\bibitem[{{Horesh} {et~al}\mbox{.}(2011){Horesh}, {Stockdale}, {Frail},
  {Kasliwal}, {Kulkarni}, {Gal-Yam}, {Arcavi}, {Ofek}, {Quimby}, {Yuan},
  {Akerlof}, {McKay}, {Weiler}, {van Dyk}, {Immler}, {Marcaide}, {Ryder},
  {Panagia}, {Pooley}, {Bauer}, \& {Williams}}]{Horesh+2011}
{Horesh} A. {et~al.}, 2011, The Astronomer's Telegram, 3411, 1

\bibitem[{{Horesh}, {Zauderer} \& {Carpenter}(2011){Horesh}, {Zauderer}, \&
  {Carpenter}}]{HoreshZC2011}
{Horesh} A., {Zauderer} A., {Carpenter} J., 2011, The Astronomer's Telegram,
  3405, 1

\bibitem[{{Iwamoto} {et~al}\mbox{.}(1997){Iwamoto}, {Young}, {Nakasato},
  {Shigeyama}, {Nomoto}, {Hachisu}, \& {Saio}}]{Iwamoto+1997}
{Iwamoto} K., {Young} T.~R., {Nakasato} N., {Shigeyama} T., {Nomoto} K.,
  {Hachisu} I., {Saio} H., 1997, \apj, 477, 865

\bibitem[{{Krauss} {et~al}\mbox{.}(2012){Krauss}, {Soderberg}, {Chomiuk},
  {Zauderer}, {Brunthaler}, {Bietenholz}, {Chevalier}, {Fransson}, \&
  {Rupen}}]{Krauss+SN2011dh-II}
{Krauss} M.~I. {et~al.}, 2012, \apjl, 750, L40

\bibitem[{{Maeda}(2012)}]{Maeda2012}
{Maeda} K., 2012, \apj, 758, 81

\bibitem[{{Maeda} {et~al}\mbox{.}(2015){Maeda}, {Hattori}, {Milisavljevic},
  {Folatelli}, {Drout}, {Kuncarayakti}, {Margutti}, {Kamble}, {Soderberg},
  {Tanaka}, {Kawabata}, {Kawabata}, {Yamanaka}, {Nomoto}, {Kim}, {Simon},
  {Phillips}, {Parrent}, {Nakaoka}, {Moriya}, {Suzuki}, {Takaki}, {Ishigaki},
  {Sakon}, {Tajitsu}, \& {Iye}}]{Maeda+2015}
{Maeda} K. {et~al.}, 2015, \apj, 807, 35

\bibitem[{{Marcaide} {et~al}\mbox{.}(1997){Marcaide}, {Alberdi}, {Ros},
  {Diamond}, {Shapiro}, {Guirado}, {Jones}, {Mantovani}, {P{\'e}rez-Torres},
  {Preston}, {Schilizzi}, {Sramek}, {Trigilio}, {Van Dyk}, {Weiler}, \&
  {Whitney}}]{Marcaide+1997}
{Marcaide} J.~M. {et~al.}, 1997, \apjl, 486, L31

\bibitem[{{Marion} {et~al}\mbox{.}(2011){Marion}, {Kirshner}, {Wheeler},
  {Vinko}, {Chornock}, {Foley}, {Challis}, {Friedman}, {Garnavich}, \&
  {Krisciunas}}]{Marion+2011}
{Marion} G.~H. {et~al.}, 2011, The Astronomer's Telegram, 3435, 1

\bibitem[{{Marion} {et~al}\mbox{.}(2014){Marion}, {Vinko}, {Kirshner}, {Foley},
  {Berlind}, {Bieryla}, {Bloom}, {Calkins}, {Challis}, {Chevalier}, {Chornock},
  {Culliton}, {Curtis}, {Esquerdo}, {Everett}, {Falco}, {France}, {Fransson},
  {Friedman}, {Garnavich}, {Leibundgut}, {Meyer}, {Smith}, {Soderberg},
  {Sollerman}, {Starr}, {Szklenar}, {Takats}, \& {Wheeler}}]{Marion+2014}
{Marion} G.~H. {et~al.}, 2014, \apj, 781, 69

\bibitem[{{Mart{\'{\i}}-Vidal}
  {et~al}\mbox{.}(2011{\natexlab{a}}){Mart{\'{\i}}-Vidal}, {Marcaide},
  {Alberdi}, {Guirado}, {P{\'e}rez-Torres}, \& {Ros}}]{Marti-Vidal+2011a}
{Mart{\'{\i}}-Vidal} I., {Marcaide} J.~M., {Alberdi} A., {Guirado} J.~C.,
  {P{\'e}rez-Torres} M.~A., {Ros} E., 2011{\natexlab{a}}, \aap, 526, A142

\bibitem[{{Mart{\'{\i}}-Vidal}
  {et~al}\mbox{.}(2011{\natexlab{b}}){Mart{\'{\i}}-Vidal}, {Marcaide},
  {Alberdi}, {Guirado}, {P{\'e}rez-Torres}, \& {Ros}}]{Marti-Vidal+2011b}
{Mart{\'{\i}}-Vidal} I., {Marcaide} J.~M., {Alberdi} A., {Guirado} J.~C.,
  {P{\'e}rez-Torres} M.~A., {Ros} E., 2011{\natexlab{b}}, \aap, 526, A143

\bibitem[{{Mart{\'{\i}}-Vidal}
  {et~al}\mbox{.}(2011{\natexlab{c}}){Mart{\'{\i}}-Vidal}, {Tudose}, {Paragi},
  {Yang}, {Marcaide}, {Guirado}, {Ros}, {Alberdi}, {P{\'e}rez-Torres}, {Argo},
  {van der Horst}, {Garrett}, {Stockdale}, \& {Weiler}}]{Marti-Vidal+2011d}
{Mart{\'{\i}}-Vidal} I. {et~al.}, 2011{\natexlab{c}}, \aap, 535, L10

\bibitem[{{Mauerhan} {et~al}\mbox{.}(2015){Mauerhan}, {Williams}, {Leonard},
  {Smith}, {Filippenko}, {Smith}, {Hoffman}, {Huk}, {Clubb}, {Silverman},
  {Cenko}, {Milne}, {Gal-Yam}, \& {Ben-Ami}}]{Mauerhan+2015}
{Mauerhan} J.~C. {et~al.}, 2015, ArXiv:1506.08844

\bibitem[{{Maund} {et~al}\mbox{.}(2011){Maund}, {Fraser}, {Ergon},
  {Pastorello}, {Smartt}, {Sollerman}, {Benetti}, {Botticella}, {Bufano},
  {Danziger}, {Kotak}, {Magill}, {Stephens}, \& {Valenti}}]{Maund+2011}
{Maund} J.~R. {et~al.}, 2011, \apjl, 739, L37

\bibitem[{{Maund} {et~al}\mbox{.}(2004){Maund}, {Smartt}, {Kudritzki},
  {Podsiadlowski}, \& {Gilmore}}]{Maund+2004}
{Maund} J.~R., {Smartt} S.~J., {Kudritzki} R.~P., {Podsiadlowski} P., {Gilmore}
  G.~F., 2004, \nat, 427, 129

\bibitem[{{Mioduszewski}, {Dwarkadas} \& {Ball}(2001){Mioduszewski},
  {Dwarkadas}, \& {Ball}}]{MioduszewskiDB2001}
{Mioduszewski} A.~J., {Dwarkadas} V.~V., {Ball} L., 2001, \apj, 562, 869

\bibitem[{{Ng} {et~al}\mbox{.}(2011){Ng}, {Potter}, {Staveley-Smith}, {Tingay},
  {Gaensler}, {Phillips}, {Tzioumis}, \& {Zanardo}}]{Ng+2011}
{Ng} C.-Y., {Potter} T.~M., {Staveley-Smith} L., {Tingay} S., {Gaensler} B.~M.,
  {Phillips} C., {Tzioumis} A.~K., {Zanardo} G., 2011, \apjl, 728, L15

\bibitem[{{Rampadarath} {et~al}\mbox{.}(2015){Rampadarath}, {Morgan}, {Soria},
  {Tingay}, {Reynolds}, {Argo}, \& {Dumas}}]{Rampadarath+2015}
{Rampadarath} H., {Morgan} J.~S., {Soria} R., {Tingay} S.~J., {Reynolds} C.,
  {Argo} M.~K., {Dumas} G., 2015, \mnras, 452, 32

\bibitem[{{Reid} \& {Brunthaler}(2004)}]{ReidB2004}
{Reid} M.~J., {Brunthaler} A., 2004, \apj, 616, 872

\bibitem[{{Roming} {et~al}\mbox{.}(2009){Roming}, {Pritchard}, {Brown},
  {Holland}, {Immler}, {Stockdale}, {Weiler}, {Panagia}, {Van Dyk},
  {Hoversten}, {Milne}, {Oates}, {Russell}, \& {Vandrevala}}]{Roming+2009}
{Roming} P.~W.~A. {et~al.}, 2009, \apjl, 704, L118

\bibitem[{{Sahu}, {Anupama} \& {Chakradhari}(2013){Sahu}, {Anupama}, \&
  {Chakradhari}}]{SahuAC2013}
{Sahu} D.~K., {Anupama} G.~C., {Chakradhari} N.~K., 2013, \mnras

\bibitem[{{Silverman}, {Filippenko} \& {Cenko}(2011){Silverman}, {Filippenko},
  \& {Cenko}}]{SilvermanFC2011}
{Silverman} J.~M., {Filippenko} A.~V., {Cenko} S.~B., 2011, The Astronomer's
  Telegram, 3398, 1

\bibitem[{{Soderberg} {et~al}\mbox{.}(2012){Soderberg}, {Margutti}, {Zauderer},
  {Krauss}, {Katz}, {Chomiuk}, {Dittmann}, {Nakar}, {Sakamoto}, {Kawai},
  {Hurley}, {Barthelmy}, {Toizumi}, {Morii}, {Chevalier}, {Gurwell},
  {Petitpas}, {Rupen}, {Alexander}, {Levesque}, {Fransson}, {Brunthaler},
  {Bietenholz}, {Chugai}, {Grindlay}, {Copete}, {Connaughton}, {Briggs},
  {Meegan}, {von Kienlin}, {Zhang}, {Rau}, {Golenetskii}, {Mazets}, \&
  {Cline}}]{Soderberg+SN2011dh-I}
{Soderberg} A.~M. {et~al.}, 2012, \apj, 752, 78

\bibitem[{{Stockdale} {et~al}\mbox{.}(2008){Stockdale}, {Weiler}, {Immler},
  {Marcaide}, {Panagia}, {Pooley}, {Sramek}, \& {van Dyk}}]{Stockdale+2008}
{Stockdale} C.~J., {Weiler} K.~W., {Immler} S., {Marcaide} J.~M., {Panagia} N.,
  {Pooley} D., {Sramek} R.~A., {van Dyk} S.~D., 2008, The Astronomer's
  Telegram, 1484, 1

\bibitem[{{Suzuki} \& {Nomoto}(1995)}]{SuzukiN1995}
{Suzuki} T., {Nomoto} K., 1995, \apj, 455, 658

\bibitem[{{Van Dyk} {et~al}\mbox{.}(2011){Van Dyk}, {Li}, {Cenko}, {Kasliwal},
  {Horesh}, {Ofek}, {Kraus}, {Silverman}, {Arcavi}, {Filippenko}, {Gal-Yam},
  {Quimby}, {Kulkarni}, {Yaron}, \& {Polishook}}]{vDyk+2011}
{Van Dyk} S.~D. {et~al.}, 2011, \apjl, 741, L28

\bibitem[{{van Dyk} {et~al}\mbox{.}(1994){van Dyk}, {Weiler}, {Panagia},
  {Rupen}, \& {Sramek}}]{vDyk+1994}
{van Dyk} S.~D., {Weiler} K.~W., {Panagia} N., {Rupen} M.~P., {Sramek} R.~A.,
  1994, \iaucirc, 5979, 2

\bibitem[{{Van Dyk} {et~al}\mbox{.}(2013){Van Dyk}, {Zheng}, {Clubb},
  {Filippenko}, {Cenko}, {Smith}, {Fox}, {Kelly}, {Shivvers}, \&
  {Ganeshalingam}}]{vDyk+2013}
{Van Dyk} S.~D. {et~al.}, 2013, \apjl, 772, L32

\end{thebibliography}

\bsp % just makes a note: ``This paper was typeset .... ``

\label{lastpage}

\end{document}